\documentclass[a4paper,12pt,notoc]{JHEP3}
\usepackage[english]{babel}
\usepackage{graphicx,epsfig}
\usepackage{cite}
\usepackage{slashed}
\usepackage{amsmath}

\newcommand{\nn}{\nonumber}
\newcommand{\be}{\begin{equation}}
\newcommand{\ee}{\end{equation}}
\newcommand{\bea}{\begin{eqnarray}}
\newcommand{\eea}{\end{eqnarray}}

\newcommand{\ba}{\begin{array}}
\newcommand{\ea}{\end{array}}
\newcommand{\balg}{\begin{align}}
\newcommand{\ealg}{\end{align}}

\newcommand{\YS}{Y^{(S)}}
\newcommand{\YSv}{Y^{(S)}_i}
\newcommand{\YSt}{Y^{(S)}_{ij}}

\newcommand{\gt}{\gamma_{ij}}
\newcommand{\raw}{\rightarrow}
\newcommand{\Hu}{{\mathcal{H}}}

\newcommand{\gsim}{\lower .75ex \hbox{$\sim$} \llap{\raise .27ex \hbox{$>$}} }
\newcommand{\lsim}{\lower .75ex \hbox{$\sim$} \llap{\raise .27ex \hbox{$<$}} }

\title{Dark Coupling and Gauge Invariance}

\author{M.B. Gavela, \\ 
Departamento de F\'{\i}sica Te\'{o}rica, Universidad
                        Aut\'{o}noma de Madrid IFT-UAM/CSIC, 28049 Cantoblanco, Madrid, Spain
}
\author{L. Lopez Honorez, \\ Departamento de F\'{\i}sica Te\'{o}rica, Universidad
                        Aut\'{o}noma de Madrid IFT-UAM/CSIC, 28049 Cantoblanco, Madrid, Spain
 and 
                        Service de Physique Th\'eorique, Universit\'e Libre de Bruxelles, Belgium}
\author{O. Mena, \\ Instituto de F\'{\i}sica Corpuscular, IFIC, CSIC and
                    Universidad de Valencia, Spain}
\author{S. Rigolin, \\ Dipartimento di Fisica, Universit\'a di Padova and  \\ 
                       INFN Padova,  Via Marzolo 8, I-35131, Padova, Italy}

\abstract{
We study a coupled dark energy--dark matter model in which the energy-momentum exchange is 
proportional to the Hubble expansion rate. The inclusion of its perturbation is required by gauge
invariance. 
We derive the linear perturbation equations for the gauge invariant energy density contrast and 
velocity of the coupled fluids, and we determine the initial conditions. The latter turn out to be 
adiabatic for dark energy,  when assuming adiabatic initial conditions for all the standard fluids. We 
perform a full Monte Carlo Markov Chain likelihood analysis of the model, using WMAP 7-year data.}

\preprint{IFT-UAM/CSIC-10-28 \\ FTUAM-10-07\\ULB-TH/10-15}
\begin{document}

\section{Introduction}
The true substance of dark energy and dark matter is unknown although it should account for 
about 95\% of the matter--energy content of our universe today~\cite{Komatsu:2010fb}. 
While the couplings of dark fluids to photons and normal matter are severely constrained~\cite{Carroll:1998zi}, 
nothing prevents dark matter--dark energy interactions~\cite{Damour:1990tw,Damour:1990eh,Wetterich:1994bg,Amendola:1999er,Zimdahl:2001ar,Farrar:2003uw,Das:2005yj,Zhang:2005jj,delCampo:2006vv,Bean:2007nx,Olivares:2007rt,Valiviita:2008iv,He:2008si,Gavela:2009cy,Jackson:2009mz,Majerotto:2009np,Valiviita:2009nu,Koyama:2009gd,Boehmer:2009tk}. At the level of the background evolution equations, it is customary to parametrize the coupling between the two dark sectors~\cite{Kodama:1985bj} as:
\begin{eqnarray}
   \label{eq:EOMm}
  \dot{\bar{\rho}}_{dm}+ 3{\mathcal{H}}\bar{\rho}_{dm} &=&a \overline{Q}_{dm}\,,\\
\label{eq:EOMe}
 \dot{\bar{\rho}}_{de}+ 3 {\mathcal{H}}\bar{\rho}_{de}(1+ w)&=&a \overline{Q}_{de}\,,
\end{eqnarray}
where  $\bar\rho_{dm}, \, \bar\rho_{de}$ denote the dark matter and dark energy energy densities, respectively,  and 
$\overline{Q}_{dm}=-\overline{Q}_{de}$ encodes the coupling between those two dark sectors and drives 
the energy exchange between them. The dot indicates derivative with respect to the conformal 
time $d\tau = dt/a$, with $\mathcal{H}={\dot a}/a \equiv a \overline{H}$ denoting the background
expansion rate, while $w \equiv w_{de} = \bar p_{de}/\bar \rho_{de}$ stands for the background dark 
energy equation of state and pressureless dark matter is assumed: $w_{dm}=\bar p_{dm}/\bar \rho_{dm}=0$. 
From now on, barred quantities are to be considered as the background quantities. 


The initial conditions for the several components populating the early universe have been explored 
to a large extent. They were first analyzed for all cosmic fluids but dark energy (see {\it e.g.}
Ref.~\cite{Ma:1995ey} and references therein), with the result that adiabatic initial conditions 
were one possibility. It was also noticed  that the  choice of gauge could be a delicate issue: 
a safe alternative proposed was to use a gauge invariant formalism \cite{Bardeen:1980kt,Kodama:1985bj,
Mukhanov:1990me,Durrer:2001gq}. The initial conditions for the case of dynamical dark energy as 
an uncoupled quintessence field have been also derived~\cite{Viana:1997mt,Dave:2002mn,
Malquarti:2002iu,Abramo:2001mv,Kawasaki:2001nx,Perrotta:1998vf,Doran:2003xq}, including a gauge invariant 
treatment~\cite{Doran:2003xq}: they turned out to be adiabatic if those for the traditional 
fluids were adiabatic. 
Furthermore, the formalism in Ref.~\cite{Doran:2003xq} has been recently applied to the case of a coupled
dark energy-dark matter systems which mimic uncoupled models at early times, both at the background and perturbation levels~\cite{Majerotto:2009np} for the viable parameter space: as expected, adiabatic initial 
conditions for dark energy naturally resulted then.
Here we consider a different class of dark couplings, not negligible at early times. 
It is also illustrated that the gauge invariant formalism is  particularly illuminating  
for the determination of the correct perturbation equations, for a general coupled theory.

The structure of the paper is as follows. In Section~\ref{sec:gauge-invar-pert}, the notation is set 
and the gauge invariant equations -at linear order in perturbation theory- for a coupled fluid are derived. 
In particular, we study in Section~\ref{sec:coupl-prop-h} the case of a (covariant) dark matter--dark energy 
interaction proportional to the Hubble rate. In Section~\ref{sec:initial-conditions}, following the method 
proposed in Ref.~\cite{Doran:2003xq}, we derive the corresponding initial conditions for dark energy. 
Then in Section~\ref{sec:data-constraints}, we constrain the type of coupled models analyzed, using several 
data sets. Section~\ref{sec:conclusion} contains the conclusions. 

\section{Gauge invariant perturbation equations}
\label{sec:gauge-invar-pert}
Following Ref.~\cite{Kodama:1985bj}, the FRW metric, up to  first order in perturbation theory, 
can be written as:
\begin{eqnarray}
g_{\mu \nu} dx^\mu dx^\nu &=& a^2 \left[- (1+2A)d\tau^2 - B_i d\tau dx^i + 
                (\gamma_{ij} +2 H_{ij})dx^i dx^j \right]\,,
\label{metric}
\end{eqnarray}
where  $\gamma_{ij}$ is the 3D flat metric with positive signature. The perturbations $A$, $B_i$ and 
$H_{ij}$ are functions of time and space and are in general gauge-dependent, {\em i.e.} not invariant 
under an infinitesimal coordinate transformation:
\begin{eqnarray}
(x^0, x^i) \rightarrow ({\hat  x}^0, {\hat  x}^i)=(x^0-T, x^i-L^i) \, . 
\end{eqnarray}
Particularizing to the case of scalar metric perturbations, two gauge invariant quantities\footnote{
The transformation properties of the metric perturbations defined in Eq.~(\ref{metric}) and the 
explicit definition of the Bardeen potentials is reminded in Appendix \ref{sec:append-gauge-invar}.} 
can be defined~\cite{Bardeen:1980kt}, the most popular being the so-called Bardeen potentials 
$\Phi_B$ and $\Psi_B$.

In describing the evolution of a given fluid ``$a$'', other gauge dependent quantities are introduced, 
such as the perturbed 4--velocity and the energy--momentum tensor, which can be expressed by:
\begin{eqnarray}
u^\mu_{a} &=&\frac{1}{a}(1-A,v^i_{a}) \,,\label{eq:umu}\\ 
T^{\mu \nu}_{a}  &=&  \bar{\rho}_{a}(1 +\delta_{a}) u^\mu_{a} u^\nu_{a} + \tau^{\mu \nu}_{a}\,,\label{eq:Tmunu}
\end{eqnarray}
where $v^i_{a}$ is the peculiar velocity perturbation of the fluid, $\delta_{a}$ the density perturbation 
and $\tau^{\mu \nu}_{a}$ the stress tensor, whose components in first order perturbation theory read 
\begin{equation}
  {\tau_{a}}^0_0 = 0\, , \qquad  {\tau_{a}}^i_0 = \bar{p}_{a}
  \, v^i_{a} \, , \qquad {\tau_{a}}^i_j = \bar{p}_{a}
  \left[\left(1+\pi^L_{a} \right)  \gamma^i_j + (\pi^T_{a})^i_j \right]\,. 
\end{equation}
In what follows, we deal with the Fourier transformations of the scalar part of the 
metric and fluid perturbations. See Appendix~\ref{sec:append-gauge-invar} for details. 
 In the equations above, $\pi^L_{a}$ and $\pi^T_{a}$ denote the isotropic and anisotropic scalar pressure 
perturbations, respectively, while $v_a$ is the scalar part of the peculiar velocity. 
Associated gauge invariant quantities can be defined, paralleling the two gauge invariant variables 
for scalar metric perturbations. Following the notation in Ref.~\cite{Doran:2003xq}, a possible gauge 
invariant formulation for $\delta_{a}$, $v_{a}$ and the stress--tensor components $\pi^L_{a}$ 
and $\pi^T_{a}$ is:
%
\begin{eqnarray}
\Delta_{a} = \delta_{a} - \frac{\dot{\bar\rho}_{a}}{\bar\rho_{a}} \frac{\mathcal{R}}{\Hu} \, ,
  && \qquad V_{a} = v_{a}- \frac{\dot H_T}{k} \label{KSv}  \\
\Gamma_{a} = \pi^L_{a} -\frac{c^2_{Aa}}{w_{a}} \delta_{a} \, , && \qquad
 \Pi_{a} = \pi^T_{a} \, \label{KSPi}.
\end{eqnarray}
The coefficient $c^2_{Aa}$ entering in the entropy perturbation $\Gamma_{a}$ is the  
adiabatic sound speed of the fluid $c_{Aa}^2 = \dot{\bar{p}}_{a}/\dot{\bar{\rho}}_{a}$ and 
$w_{a}$ is the equation of state of the fluid.

We focus next on the derivation of the gauge invariant equations for the matter density contrast 
$\Delta_{a}$ and the fluid velocity $V_{a}$, for a generic coupled fluid.

\subsection{Coupled fluids in general}
\label{sec:general-case}
Consider the full (background plus perturbations) continuity equation for  fluid ``$a$'':
\begin{equation}
  \nabla_\mu T^{\mu\nu}_{a} = Q^\nu_{a} \qquad , \qquad \sum_a Q^\nu_{a}=0\,,
\label{eq:conserv}
\end{equation}
where $T^{\mu\nu}_{a}$ denotes the  corresponding energy-momentum tensor and the vector $Q^\nu_{a}$ governs 
the energy-momentum transfer. The constraint on the right accounts for total energy--momentum 
conservation. Following Ref.~\cite{Kodama:1985bj}, $Q^\nu_{a}$ can be written as:
\begin{eqnarray}
Q^\mu_{a} & = & Q_{a} u_{a}^\mu + j^\mu_{a}\,, \qquad \quad {\rm with} \qquad \quad 
                  j^\mu_{a} u_\mu^{a} =0\,, \\
Q_{a} &=& \overline{Q}_{a} \left( 1 + \frac{\delta Q_{a}}{\overline{Q}_{a}} \right) \equiv
                   \overline{Q}_{a} \left( 1 + \varepsilon_{a} \right)\,,
\end{eqnarray}
where $j^\mu_{a}$  and $\varepsilon_{a}$ are perturbation parameters. In particular, the background 
contributions  reduce to the coupled dark energy-dark matter case in Eqs.~(\ref{eq:EOMm}) and~(\ref{eq:EOMe}), for $Q^\nu_{de}=-Q^\nu_{dm}$.
Defining for simplicity $j^i_{a}=\bar \rho_{a} f^i_{a}/a$, the total coupling reads 
\begin{eqnarray}
Q^\mu_{a} = \frac{1}{a}\left(\overline{Q}_{a} \left[1-\left(A -{\varepsilon}_{a} \right)\right], 
\overline{Q}_{a} v^i_{a} + \bar\rho_{a} f^i_{a}\right) \, .\label{coup}
\end{eqnarray} 
Let's denote by $f_{a}$ the Fourier transform of the scalar part of $f_{a}^i$. One can show that 
$f_{a}$ is invariant under gauge transformations, while ${\varepsilon}_{a}$ transforms as
\begin{equation}
\widehat{\varepsilon}_{a} = \varepsilon_{a}-\frac{\dot{\overline{Q}}_{a}}{\overline{Q}_{a}} T\,,
\label{epsgauge}
\end{equation}
where the ``hat" denotes gauge transformed quantities. This suggest a possible choice of gauge invariant 
variables for the coupling perturbation parameters, given by 
\begin{eqnarray}
E_{a} &=& \varepsilon_{a}-\frac{\dot{\overline{Q}}_{a}}{\overline{Q}_{a}}
            \frac{\cal R}{\cal H }\,,  \label{epsGI}\\
F_{a} &=& f_{a} \,.
\end{eqnarray}
The gauge invariant choice in Eq.~(\ref{epsGI}) is analogous to that for $\Delta_{a}$ in Eq.~(\ref{KSv}).

With the help of these variables, the scalar perturbation equations for the matter density contrast
$\Delta_{a}$ and the peculiar velocity $ V_{a}$, for a generic coupled fluid, read:
\begin{eqnarray}
\hspace{-0.5cm} 
\dot\Delta_{a}  &=& 
    -3\Hu\left[ \left(c^2_{Aa} - w_{a}\right) \Delta_{a} + w_{a} \Gamma_{a} \right]-k(1+w_{a})V_{a} 
     +3\Hu \bar q_{a}\left[ \mathcal{A} + E_{a}-\Delta_{a} \right]~,    \label{ddotGIcW}  \\  
\dot V_{a}  &=& 
    -\Hu\left(1-3c^2_{Aa}\right) V_{a} +
    \frac{k}{1+w_{a}}\left[c^2_{Aa} \Delta_{a} + w_{a}\left(\Gamma_{a} -
    \frac{2}{3} \Pi_{a} \right) \right] + k\left(\Psi_B -3c^2_{Aa}\Phi_B\right) \nn \\
  &&  -3\Hu \bar q_{a} \frac{c^2_{Aa}}{1+w_{a}} 
    \left(V_{a} -  k \frac{\Phi_B}{\Hu}\right) + \frac{a  F_{a}}{1+w_{a}}~, \label{vdotGIcW}  
\end{eqnarray}
where $\mathcal{A}$ is a metric gauge invariant quantity~\cite{Kodama:1985bj}, whose expression 
is given in Eq.~(\ref{eq:KSA}).
The quantity $\bar q_{a}$ accounts for the energy transfer $\overline Q_{a}$ in Eqs.~(\ref{ddotGIcW})  
and (\ref{vdotGIcW}), rescaled as follows 
\begin{eqnarray}
\bar q_{a} \equiv \frac{a \overline{Q}_{a} }{3 \Hu \bar{\rho}_{a}} \, .
\end{eqnarray}
For vanishing $\bar q_{a}$ and $F_a$, Eqs.~(\ref{ddotGIcW})  and (\ref{vdotGIcW}) reduce to those in Ref.~\cite{Doran:2003xq}.

\subsection{Coupling proportional to $H$}
\label{sec:coupl-prop-h}

Coupled models with a dark matter-dark energy coupling proportional to the Hubble expansion rate   
have been studied at the level of linear perturbations in several recent works, see for example 
Refs.~\cite{Valiviita:2008iv,He:2008si,CalderaCabral:2008bx,Gavela:2009cy,Jackson:2009mz}. 
Perturbations in the expansion rate were neglected, though. To analyze the issue, the results 
of the previous section will be particularized to the following coupling:   
\begin{equation}
Q^\nu_{dm}= \xi H \rho_{de} \, u^{\nu}_{dm}=-Q^\nu_{de}\,.
\label{eq:coupl}
\end{equation}
Here $Q^\nu_{dm}$ is chosen parallel to the dark matter four velocity $u^{\nu}_{dm}$ to avoid 
momentum transfer in the dark matter rest frame~\cite{Valiviita:2008iv}. 
The  evolution equation for the dark matter velocity remains then equal to that of baryons, avoiding the
violation of the weak equivalence principle. Moreover, the authors of Ref.~\cite{Valiviita:2008iv} pointed 
out that such a coupled models could suffer from non-adiabatic instabilities if the coupling $Q_{dm}$ 
is chosen proportional to the dark matter energy density. In Ref.~\cite{He:2008si, Gavela:2009cy, 
Jackson:2009mz}, it was shown though that such instabilities could be avoided in a minimal way choosing 
a coupling $Q_{dm}$ proportional to the dark energy density\footnote{Would an interaction proportional 
to the dark matter density be studied instead,  it would be necessary to consider a time dependent dark 
energy equation  of state, in order to avoid early time instabilities, thus introducing at least one extra free 
parameter, see Ref.~\cite{Majerotto:2009np}.}. 

It is important to notice that, in order to deal with a consistent model, $H$ in Eq.~(\ref{eq:coupl}) 
must denote the total expansion rate (background plus perturbations), $H=\overline{H}+\delta H$, 
while in all previous studies only the background quantity was considered. The inclusion of $\delta H$ 
is mandatory to preserve gauge invariance, as we proceed to illustrate. For the model in Eq.~(\ref{eq:coupl}) 
one obtains:
\begin{eqnarray}
  \overline{Q}_{dm} &=& \xi \overline{H} \bar{\rho}_{de}\,,
  \label{ourcupback}\\
   \varepsilon_{dm} &=& \frac{\delta H}{H} + \frac{\delta \rho_{de}}{\bar \rho_{de}} \equiv 
   {\cal K }  + \delta_{de}\,.
\end{eqnarray}
The ${\cal K}$ term (see Eq.~(\ref{eq:kappam})), represents the expansion rate perturbation, overlooked in all the references mentioned above. Indeed, ${\cal K}$ depends on the time slicing, 
so that the coupling perturbation $\varepsilon_{a}$ gauge transforms as:
\begin{eqnarray}
\widehat\varepsilon_{a}-\varepsilon_{a} & \equiv & \frac{\dot{\overline{Q}}_{a}}{\overline{Q}_{a}} T 
     = \frac{\dot{\overline{H}}}{\overline{H}} \, T + 
       \frac{\dot{\overline{\rho}}_{de}} {\overline{\rho}_{de}}  \, T 
   \; = \; \left(\widehat{\mathcal{K}} - {\mathcal{K}}
   \right)+\left(\widehat \delta_{de}-\delta_{de}\right).
\label{eq:epsvar}
\end{eqnarray}
To our knowledge this result was not explicitly discussed elsewhere. We will see in 
Sec.~\ref{sec:data-constraints} that the extra contribution resulting from $\delta H$ has little 
quantitative impact on the physical constraints obtained from data, while being essential for gauge invariance.

Before proceeding further let us comment on the covariance of the coupling of Eq.~(\ref{eq:coupl}). 
First of all, the dark energy density can be rewritten as $\rho_{de}= T^{\mu\nu}_{de} u^{de}_\mu 
u^{de}_\nu$. 
Moreover, we can express the Hubble expansion rate in terms of the
covariant derivative of the four velocity defined in 
Eq.~(\ref{eq:umu}). Indeed, it is straightforward to verify that the
background quantity associated to $u^\mu_{a;\mu}$
is directly proportional to the expansion rate ${\overline H}$. Following~\cite{Kodama:1985bj} one has: 
\begin{equation}
  \Theta_a= u^\mu_{a;\mu}= 3\overline H(1+ \mathcal{K}_a)\, . \label{eq:Thetaa}
\end{equation}
Under gauge transformations, the perturbation $\mathcal{K}_a$ (associated to the {\it a}--fluid)
transforms like:
\begin{equation}
\widehat{\mathcal{K}_a} - {\mathcal{K}_a}=\frac{\dot{\overline{H}}}{\overline{H}} \, T 
\end{equation}
which is exactly what is needed to preserve the gauge invariance of the coupled model under study, 
see Eq.~(\ref{eq:epsvar}). In the following,  we will use for definiteness the total matter expansion 
rate $ \Theta_T=u^\mu_{T;\mu}= 3\overline H(1+ \mathcal{K})$, denoting with $\cal K$ the 
perturbation associated to the total fluid. Finally the coupling of Eq.~(\ref{eq:coupl}) 
can be written in a covariant way as: 
\begin{equation}
Q^\nu_{dm}= \xi \frac{\Theta_T}{3}\, T^{\alpha\beta}_{de} u^{de}_\alpha u^{de}_\beta \, u^{\nu}_{dm}=-Q^\nu_{de}\,.
\label{eq:couplcov}
\end{equation}


We can now particularize Eqs.~(\ref{ddotGIcW}) and (\ref{vdotGIcW}) to our coupling. 
Expressing $\cal K$ in terms of gauge invariant quantities one obtains:
\bea
E_a = \Delta_{de} + \left( \frac{x^2}{3} -\frac32 (1+w_T)  \right)\widetilde{V}_T + 2 \Phi_B
\label{newEa}
\eea
where $w_T$ and $V_T$ is the equation of state and velocity of the total fluid.
The density and velocity perturbation equations then read:
\begin{eqnarray}
&& \hspace*{-1cm}\frac{\dot{\Delta}_{dm}}{\Hu} = - x^2\, \widetilde{V}_{dm} + \xi \frac{\bar{\rho}_{de}}{\bar{\rho}_{dm}}
       \left[ \left( \Delta_{de} - \Delta_{dm}\right) + \frac{x^2}{3}\widetilde{V}_T\right]~, \label{ourfinalddm}\\
&& \hspace*{-1cm}\frac{\dot{\widetilde{V}}_{dm}}{\Hu} = - \left( 1- \frac{\dot\Hu}{\Hu^2}\right)\widetilde{V}_{dm} - \left( \Phi_B + \Omega_{\nu} \widetilde{\Pi}_{\nu} \right)~,
        \label{ourfinaltdm}\\
&& \hspace{-1cm}
\frac{\dot{\Delta}_{de}}{\Hu} =  - 3 (c^2_{S}-w) \Delta_{de} - (1+w)\, x^2\, \widetilde{V}_{de}  
     +9\left(1 + w \right)\left(c^2_S -c^2_A \right) \left(\Phi_B - \widetilde{V}_{de} \right)  \nn \\
&&  \hspace{0.3cm}
    -\xi  \left[  \frac{x^2}{3}\widetilde{V}_T-3\left(c^2_S -c^2_A \right)\left(\Phi_B -\widetilde{V}_{de}\right)\right]~,
\label{ourfinaldde}\\
&& \hspace{-1cm} 
\frac{\dot{\widetilde{V}}_{de}}{\Hu} = - \left(1- \frac{\dot\Hu}{\Hu^2}-3 c^2_S\right) \widetilde{V}_{de} -
    \left( 1 +3 c^2_S\right)\Phi_B - \Omega_{\nu} \widetilde{\Pi}_{\nu} 
    +\frac{c^2_S}{1+w} \Delta_{de} + \nn \\
&& \hspace{0.3cm} +\frac{\xi}{1+w}  \left[ \left(1+c^2_S\right) \widetilde{V}_{de} - \widetilde{V}_{dm} 
    - c^2_S\Phi_B \right]~, 
\label{ourfinalvde}
\end{eqnarray}
the rescaled quantities $\widetilde{V}=V/x$ and $\widetilde{\Pi}=\Pi/x^2$ were used, with $x=k/\Hu$. 
In deriving these equations, the dark energy entropy perturbation $\Gamma_{de}$ has been rewritten in terms 
of $\Delta_{de}, V_{de}$ and $\Phi_B$, see Eq.~(\ref{eq:Gammac2}). 
$c^2_A$ and $c^2_S$ are the dark energy adiabatic sound speed and the rest frame sound speed, respectively. 
In the following we work in the framework of constant $w$, $c_A^2=w$ and $c^2_S=1$. 


%
\section{Initial conditions}
\label{sec:initial-conditions}
In Ref.~\cite{Doran:2003xq}, whose gauge invariant formalism we follow, the solution of the system of differential 
equations for the perturbations is reduced to that of a simple eigenvalues/eigenvectors problem:
\begin{eqnarray}
 U' \equiv \frac{d U}{d\ln x} = A(x)\, U\,.
\label{eq:syst}
\end{eqnarray}
Here $A(x)$  encodes the evolution equations for all the universe components, and
\begin{equation}
 U^T\equiv\{\Delta_{dm},\tilde V_{dm}, \Delta_{\gamma},\tilde V_{\gamma},
  \Delta_{b},\Delta_{\nu},\tilde
  V_{\nu},\tilde\Pi_{\nu},\Delta_{de},\tilde V_{de}\} 
\end{equation}
is an array of gauge invariant perturbations, where the subscripts $\gamma$, $b$ and $\nu$ stand 
for photons, baryons and neutrinos, respectively.
No anisotropic stress for dark energy and negligible anisotropic stress for photons (due to large 
Thompson damping) are assumed. 

The evolution equations for baryons, photons, and neutrinos are unaltered by the presence of the dark coupling 
and we obviate them below. 
In contrast, the dark matter and dark energy perturbation equations for the case under study are significantly modified. 
The exact form of the correspondent $A(x)$ matrix can be easily derived from Eqs.~(\ref{ourfinalddm})--(\ref{ourfinalvde}). 

To obtain the initial conditions for cosmological perturbations, it is necessary to study the  evolution 
of the several cosmic components at a very early stage, when the universe was radiation dominated and 
$\Hu = 1/\tau$. One is interested in the time dependence of all perturbations on super--horizon 
scales, {\it i.e} for $x = k\tau \ll 1$.  

\subsection{The $A_0$ matrix}
\label{sec:ax-matrix}
%

At early times $x\ll 1$, the $A(x)$ matrix can be approached by a constant matrix $A_0$, if no 
divergence appears when taking the $\lim_{x\rightarrow 0} A(x)$. The assumption that the universe 
is radiation dominated at early times implies $w_T=1/3$, $\bar\rho_T=\bar\rho_{rad}$ and 
\begin{eqnarray}
  \Omega_\nu =\bar\rho_\nu/\bar  \rho_{rad} = R_\nu,  \quad \Omega_\gamma =
  1- R_\nu\quad \mbox{and}\quad\frac{\Omega_{de}}{ \Omega_{dm}}=\frac{\bar\rho_{de}}{ \bar\rho_{dm}} \propto x^{-(3w+\xi)}\,.
\end{eqnarray}
 $w<-1/3$ is assumed as well, in order to obtain cosmic acceleration, which implies that $(3w+\xi)$ can be 
always taken negative for $\xi<0$. 

Using Eqs.~(\ref{ourfinalddm})--(\ref{ourfinalvde}) and taking the 
$x\rightarrow 0$ limit, the following entries in the $A_0$ matrix associated to 
$\Delta_{de}$ and $\tilde V_{de}$ result:
{\footnotesize
\vspace{0.5cm}
\bea
\hspace{-0.35cm}{
  \label{eq:A0}
\left(
\begin{array}{cccccccccc}
 0 & 0 & \frac{R_\gamma}{4} (\alpha+\beta\xi) &R_\gamma (\alpha+\beta\xi) & 0 & 
      \frac{R_\nu}{4} (\alpha+\beta\xi)& R_\nu(\alpha+\beta\xi)  & 0 & -\beta &-(\alpha+\beta\xi) \\
&&&&&&&&&\\ 
0 & -\xi_r &- R_\gamma \left(1+\xi_r/4\right) & -4 R_\gamma \left(1+\xi_r/4\right)  & 0 & 
   -R_\nu\left(1+\xi_r/4 \right) &-4R_\nu \left(1+\xi_r/4\right)& -R_\nu & \frac{1}{1 + w} & 1+2\xi_r
\end{array} \right)}\nn \\
& \nn 
\eea
}
where $\alpha=9(1-w^2)$, $\beta=3(1-w)$, $\xi_r=\xi/(1+w)$ and $R_\gamma=1 - R_\nu$. 
The other lines in the $A_0$ matrix remain equal to the uncoupled case ones.
Indeed the extra term in the $\Delta_{dm}$ equation proportional to $\xi \bar\rho_{de}/\bar \rho_{dm}$ 
can be safely neglected in the $x\rightarrow 0$ approximation. We thus recover the standard non 
interacting dark matter perturbation equation in the early universe. Notice that this was also the case 
of the viable coupled model discussed in Ref.~\cite{Majerotto:2009np}.

\subsection{Adiabatic initial conditions }
\label{sec:adiab-init-cond}
Let $U_i$ be an eigenvector of $A_0$ with eigenvalue $\lambda_i $. The solution to the system in 
Eq.~(\ref{eq:syst}) can then be expressed as a linear combination of $x^{\lambda_i}U_i$ terms. 
Those corresponding to the largest eigenvalues will dominate the time evolution. We checked that 
for the model under study, Eq.~(\ref{eq:couplcov}), the dominant modes are associated to $\lambda_i=0$ 
values and they suffice to specify the initial conditions. The subdominant modes decay in time as they 
correspond to negative real eigenvalues\footnote{Also, see Ref.~\cite{Doran:2003xq} for the case of 
quintessence and Ref.~\cite{Majerotto:2009np} for coupled dark sectors with a different coupling.}.
The  dominant eigenvalue of the evolution matrix, $\lambda_i=0$, is fourfold degenerate (as was the 
case for a universe without dynamical dark energy) and the corresponding four eigenvectors serve as a 
convenient basis to specify the initial conditions. 

Let us assume adiabatic initial conditions for all species but dark energy, as strongly constrained 
by WMAP data~\cite{Komatsu:2010fb,Larson:2010gs}. For each pair of components $a_1$ and $a_2$, the relative entropy 
perturbation, $S_{a_1 a_2}$, vanishes:  
\begin{equation}
 S_{a_1 a_2} = \frac{\Delta^{0}_{a_1}}{\dot{\bar\rho}_{a_1}/\bar\rho_{a_1}} - 
          \frac{\Delta^{0}_{a_2}}{\dot{\bar\rho}_{a_2}/\bar\rho_{a_2}} =0~.
\label{eq:ad}
\end{equation}
For baryons, neutrinos, photons and dark matter this implies: 
\begin{equation}
\Delta^{0}_{dm}=\Delta^{0}_{b}=\frac34 \Delta^{0}_{\gamma}= \frac34 \Delta^{0}_{\nu}~,
\label{eq:ad-standard}
\end{equation}
from which one obtains:
\begin{eqnarray}
\widetilde V^{0}_{\gamma}= \widetilde V^{0}_{b}=\widetilde V^{0}_{\nu}= \widetilde V^{0}_{dm}&=& 
-\frac{5}{4}\,{\cal P} \Delta_\gamma^0 \qquad \mbox{and} 
\qquad\widetilde \Pi_\nu^{0}=-{\cal P}\Delta_\gamma^0~,
\label{eq:ad-v}
\end{eqnarray}
with ${\cal P}= 1/(15+4 R_\nu)$.
Those are the standard adiabatic initial conditions for velocity perturbations and anisotropic stress. 
 Solving the eigenvalue problem for our $A_0$ matrix, it follows that dark energy also obeys adiabatic initial conditions given by 
   \begin{eqnarray}
 \Delta^{0}_{de}&=&\frac{3}{4}\left(1+w+\frac{\xi}{3}\right)\Delta^{0}_{\gamma}\label{eq:deltade0} \, ,\\
  \widetilde{V}^{0}_{de}&=&-\frac{5 \cal P}{4}\Delta_\gamma^0\label{eq:Vde0}\,.
\end{eqnarray}
Consequently, adiabatic initial conditions for the matter and radiation components automatically 
imply adiabatic initial conditions for dark energy, alike to the case  for tracking scalar quintessence~\cite{Doran:2003xq} or those obtained for dark energy-dark matter couplings which 
do not depend explicitly on the Hubble rate\footnote{See Ref.~\cite{Majerotto:2009np} for $Q_a=
\pm \Gamma\rho_{dm}$, where $\Gamma$ is a constant.}. 

As a final comment, notice that the previous results do not depend on the fact that we are using 
the expansion of the total fluid $\Theta_T$ (and its perturbation $\cal K$) to define the dark
coupling in Eq.~(\ref{eq:couplcov}). In fact one could have used the expansion rate of any single 
specie, $\Theta_a$.
In that case, Eq.~(\ref{newEa}) 
should be replaced by: 
\bea
E_a = \Delta_{de} +  \frac{x^2}{3} \widetilde{V}_a-\frac32 (1+w_T) \widetilde{V}_T + 2 \Phi_B
\eea
which implies that all the contributions of the dark coupling going as $x^2 \widetilde{V}_T$ in 
Eqs.~(\ref{ourfinalddm}) and (\ref{ourfinaldde}) should be replaced with $x^2 \widetilde{V}_a$.
This does not modify the expression of the matrix $A_0$ of Sec.~\ref{sec:ax-matrix}, that encodes 
the evolution equations at early times, as in the limit $x\rightarrow 0$ all the $x^2$--terms can 
be neglected. As a consequence our results of Eqs.~(\ref{eq:deltade0}), and~(\ref{eq:Vde0}) would 
not be affected, and our conclusion on adiabatic initial conditions remains unchanged.

\section{Data constraints}
\label{sec:data-constraints}
%
\begin{figure}[t]
\vspace{-1cm}
\begin{center}
\begin{tabular}{cl}
\hspace*{-0.75cm} 
\includegraphics[width=7.5cm]{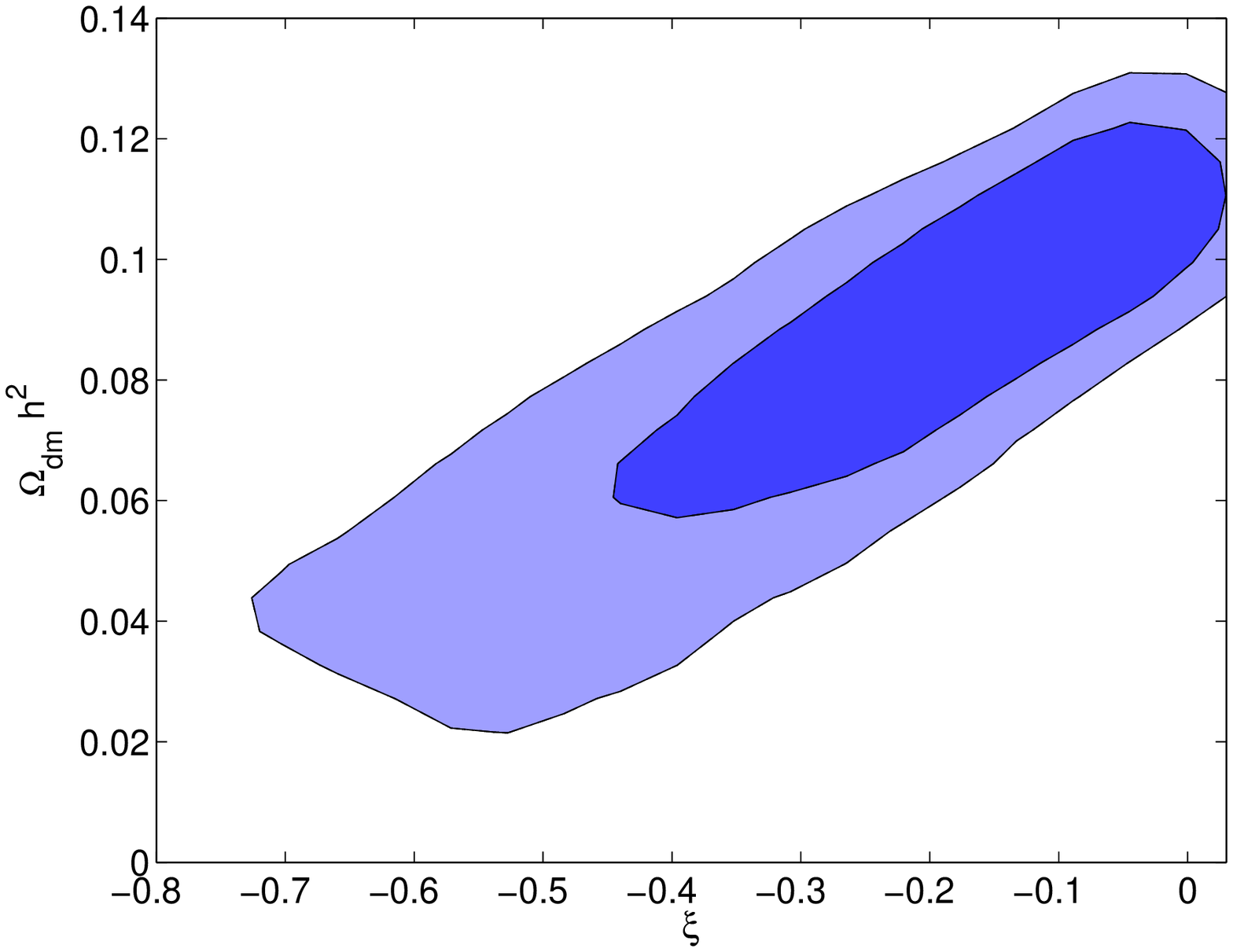} & 
\includegraphics[width=7.5cm]{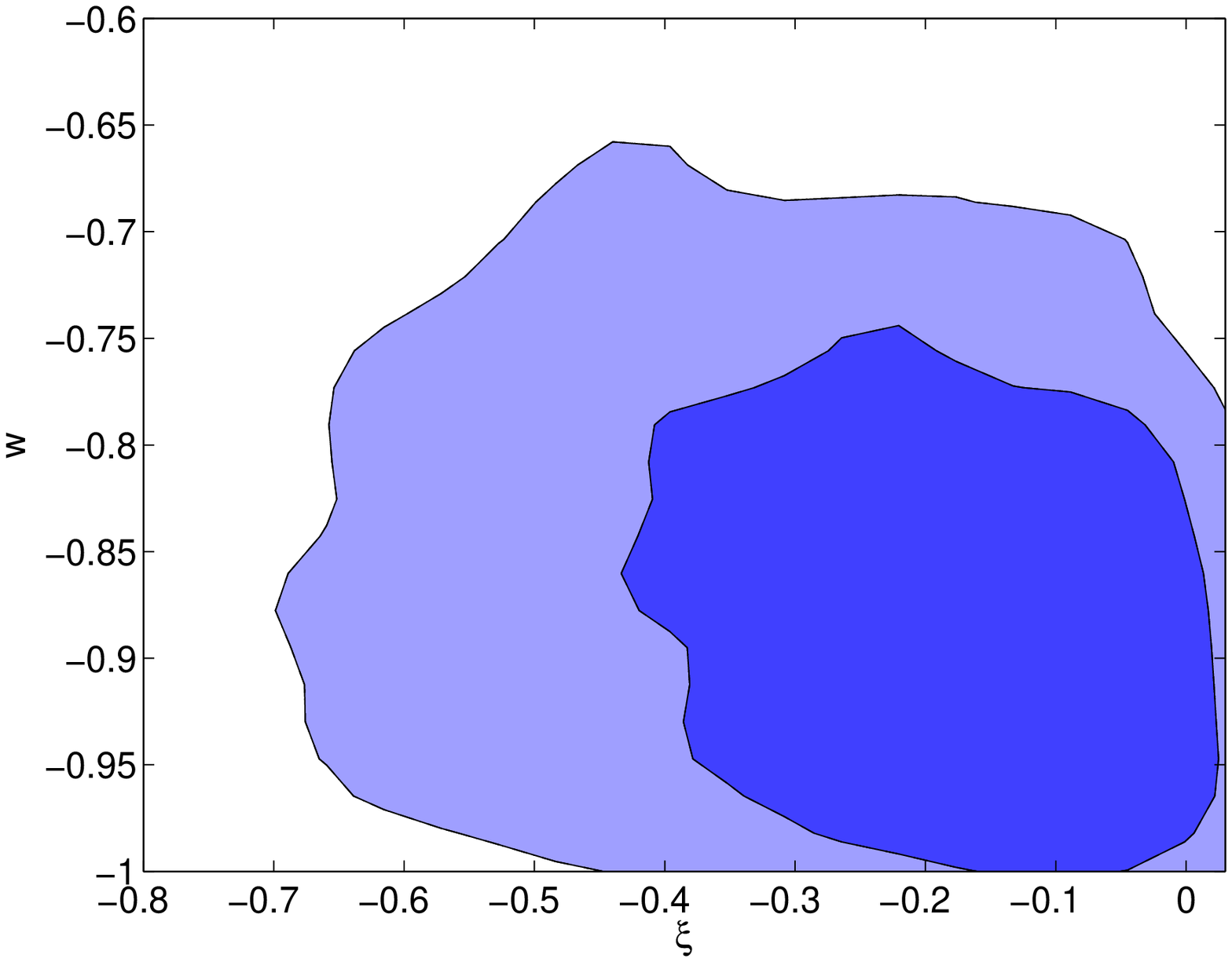} 
\end{tabular}
\caption{
\it Left (right) panel: 1$\sigma$ and 2$\sigma$ marginalized contours in the $\xi$--$\Omega_{dm} h^2$ 
($\xi$--w) plane. The contours show the current constraints from WMAP7, HST, SN, $H(z)$ and LSS data 
taking into account the expansion rate perturbation ${\mathcal{K}} $.}
\label{fig:fig0o}
\end{center}
\vspace{-0.1cm}
\end{figure}

In this section we briefly revisit the constraints on the dark coupling $\xi$ presented in 
Ref.~\cite{Gavela:2009cy}, adding to the analysis the contribution from the expansion rate
perturbation ${\mathcal{K}}$ and imposing adiabatic initial conditions for all fluids. We 
have therefore modified the Boltzmann CAMB code~\cite{Lewis:1999bs} to incorporate the dark 
coupling $\xi$ and the ${\cal K} $ terms. 

In the synchronous gauge, ${\mathcal{K}}= \theta_T/(3\Hu) + \dot h/(6\Hu)$ and the perturbation 
equations reduce to:
 \begin{eqnarray}
\label{eq:deltambe}
\dot\delta_{dm}  & = & -(k v_{dm}+\frac12 \dot h)
+\xi {\mathcal H}\frac{\rho_{de}}{\rho_{dm}} \left(\delta_{de}-\delta_{dm}\right)+\xi \frac{\rho_{de}}{\rho_{dm}} \left(\frac{k v_T}{3}+\frac{\dot h}{6}\right)\,, \\
 \label{eq:thetames}
 \dot v_{dm}  & = & -{\mathcal H} v_{dm} \,,\\
\label{eq:deltaees}
\dot\delta_{de}  & = & -(1+w)(k v_{de}+\frac12 \dot h)
   -3 {\mathcal H}\left(1 -w\right)\left[ \delta_{de} +
   {\mathcal H} \left( 3(1+w) + \xi\right)\frac{v_{de}}{k} \right]\,
 \\ \nonumber
&& -\xi \left(\frac{k v_T}{3}+\frac{\dot h}{6}\right) \,,\\
\label{eq:thetaees}
\dot v_{de}  & = & 2 {\mathcal H}\left(1 +\frac{\xi}{1+w} \right)
    v_{de}+\frac{k}{1+w}\delta_{de}-\xi{\cal H}\frac{v_{dm}}{1+w}\,,
\end{eqnarray}
where $v_T$ is defined in Eq.~(\ref{eq:vt}).

We have extracted the cosmological parameters by means of the publicly available Markov Chain Monte Carlo 
package \texttt{cosmomc}~\cite{Lewis:2002ah}. The cosmological model is described by ten free parameters
\begin{equation}
\left\{\omega_{b}, \omega_{dm}, \theta_{CMB}, \tau, \Omega_k, f_\nu, w, \xi, n_s, A_s \right\}~, \nn
\end{equation}
where $\omega_{b}=\Omega_{b} h^2$ and $\omega_{dm}=\Omega_{dm} h^2$ are the current baryon and 
dark matter densities respectively, $\theta_{CMB}$ is proportional to the ratio of the sound 
horizon to the angular diameter distance, $\tau$ is the reionization optical depth, $\Omega_k$ 
is the spatial curvature, $f_\nu = \Omega_\nu/\Omega_{dm}$ refers to  the neutrino fraction, 
$n_s$ is the scalar spectral index and $A_s$ the amplitude of the primordial spectrum. 

The analysis is restricted  to negative couplings and  also $w>-1$ (to ensure the avoidance of phantom 
behaviour), exactly as it we did previously in Ref.~\cite{Gavela:2009cy}.
The basic data set we exploit here includes a prior on the Hubble parameter of $72\pm8$~km/s/Mpc 
from the Hubble key project (HST)~\cite{Freedman:2000cf}, the constraints coming 
from the latest compilation of supernovae (SN)~\cite{Kowalski:2008ez}, the matter power spectrum 
(large scale structure data or LSS data) from the spectroscopic survey of Luminous Red Galaxies 
from the Sloan Digital Sky Survey survey~\cite{Tegmark:2006az}, the $H(z)$ data from galaxy  ages~\cite{Simon:2004tf} and the WMAP7 data~\cite{Komatsu:2010fb,Larson:2010gs}. 

CMB constraints the amount of dark matter at redshift $\sim 1000$.
In the presence of a negative dark coupling, the energy flows from dark matter to dark energy, 
thus dark matter energy density is smaller today as it can be seen in
Fig.~\ref{fig:fig0o} (left panel). This effect is compensated for
large scale structures by a larger growth of dark matter perturbation (see {\it
e.g.}~\cite{CalderaCabral:2009ja}). 
Figure~\ref{fig:fig0o}, left (right) panel illustrates the $1\sigma$ and $2\sigma$ marginalized contours 
obtained in the $\xi$--$\Omega_{dm} h^2$ ($\xi$--$w$) plane.  We verified that the results do not differ significantly if including WMAP5 data (as we had done in Ref.~\cite{Gavela:2009cy}) instead of WMAP7 data. 

Overall, the results show that the addition to the analysis of the perturbation expansion rate 
${\mathcal{K}}$ leaves basically unaffected the quantitative constraints on the cosmological parameters previously obtained in Ref.~\cite{Gavela:2009cy}. Indeed, all the additional terms introduced to make perturbations gauge invariant give negligible contributions at observable scales.

\section{Conclusions}
\label{sec:conclusion}

Interacting dark energy-dark matter cosmologies in which the coupling
term is proportional to the Hubble  expansion rate are revisited. 
While in previous works the perturbation in the Hubble expansion rate
was neglected, it is illustrated here  how the inclusion of such a
term is mandatory to satisfy the gauge invariance of the theory. It
also serves as a guide to define a covariant formulation of the dark sector
interaction. In this work, the latter has been chosen to be expressed
in terms of the expansion rate associated to the total fluid. This
choice is however not unique, we could have used the expansion rate
of any other fluid.   
  For the case under study, 
we compute the linear perturbation evolution  using 
a gauge invariant formalism. After imposing adiabatic initial conditions on the matter and radiation fluids, we find that
the initial conditions for the coupled dark energy fluid are also
adiabatic. This result is independent of the choice in the covariant
formulation of the expansion rate.  
The new terms arising from the expansion rate 
perturbation have negligible quantitative impact on the constraints on cosmological parameters previously obtained in the literature.  A new analysis has been performed using the latest WMAP7 data.

\section*{Acknowledgments}
B.~G. and L.~L.~H  are supported by CICYT through the project
FPA2009-09017 and by CAM through the project 
HEPHACOS, P-ESP-00346. L.~L.~H. ackowledges the partial support of the F.N.R.S. and  the I.I.S.N.. 
O.~M. work is supported by the MICINN Ram\'on y Cajal contract, AYA2008-03531 and CSD2007-00060.
S.~R. acknowledges the partial support of an Excellence Grant of Fondazione Cariparo and of the European
Program ``Unification in the LHC era'' under the contract
PITN-GA-2009-237920 (UNILHC). All the authors acknowledge partial
support by the PAU (Physics of the accelerating universe) Consolider Ingenio 2010.

\appendix
\section{Gauge invariant formalism }
\label{sec:append-gauge-invar}
The conventions we use are mostly from  Ref.~\cite{Kodama:1985bj} with a few exceptions. For perturbations 
in flat space time, the perturbation variables can be expanded by harmonic functions $Y^{(S)}(x,k)$ 
satisfying to $(\nabla_x+k^2)Y^{(S)}=0$.
In the following we focus on scalar perturbations for which we define:
\begin{eqnarray}
Y^{(S)} _i & = & - \frac{1}{k} Y^{(S)}_{|i}~,  \\
Y^{(S)} _{ij} & = & \frac{1}{k^2} Y^{(S)}_{|ij} + \frac{1}{3}
\gamma_{ij} Y^{(S)}~.  
\end{eqnarray}
\subsection{Metric perturbations}
For the metric defined in Eq.~(\ref{metric}),  expanding in the Fourier basis the independent perturbations, 
we denote:
\begin{eqnarray}
A        &\raw & \widetilde{A} \YS~, \nn \\
B_i     & \raw & \widetilde{B}_L \YSv~,  \nn \\
H_{ij}  & \raw & \widetilde{H}_L \gt + \widetilde{H}_T \YSt~, \nn 
\end{eqnarray}
where $\widetilde{H}_{ij} \gamma^{ij}=0$. From now on, for sake of simplicity we will drop the tilde symbols. Remember that all these quantities 
are represented by the correspondent Fourier expansion and depend only on time and on the 
3-momentum $k$, while the position dependence is left only in the basis $Y$ elements.

Gauge transformations are associated to infinitesimal coordinate transformations under which:
$(x^0, x^i) \rightarrow ({\hat  x}^0, {\hat  x}^i)=(x^0-T, x^i-L^i)$. 
It can be shown that the metric perturbation transforms as:
\begin{eqnarray}
\widehat A -A &=& \Hu T + \dot{T}~, \\
\widehat B -B &=& - k T - \dot{L}~, \\
\widehat H_L - H_L &=& H_L +  kL/3 + \Hu T~, \\
\widehat H_T - H_T &=& H_T -k L~. 
\end{eqnarray}

Before going to the gauge invariant variable definition, let us define some useful metric quantities 
and their transformations:
\begin{eqnarray}
\sigma_g &=&\frac{1}{k}\left(\dot{H}_T - k B\right)~,  \label{eq:sg}\\
{\mathcal{R}} &=& H_L + \frac{1}{3} H_T~, \label{eq:R}\\ 
{\mathcal{K}}   &=& \frac{1}{\Hu} \left[ -\Hu A + \frac{k}{3} v_T + \dot{H}_L \right]~,  
\label{eq:kappam}
\end{eqnarray}
where $v_T$ is the center of mass velocity for the total fluid, satisfying
\begin{equation}
  (1+w_T) v_T=\sum_a (1+w_a)\Omega_a v_a \,. \label{eq:vt}
\end{equation}
In the text is also sometimes used the following quantity: 
\bea
{\mathcal{K}}_a   &=& \frac{1}{\Hu} \left[ -\Hu A + \frac{k}{3} v_a + \dot{H}_L \right]\, . 
\eea
The physical meaning of the quantities above is the following: $\sigma_g$ represents the shear perturbation, 
${\mathcal{R}}$ is the curvature perturbation and ${\mathcal{K}}$ (${\mathcal{K}}_a$) is the expansion rate perturbation of the total ({\it a}) fluid. These quantities are not gauge invariant but transform as:
\begin{eqnarray}
\widehat \sigma_g - \sigma_g &=&  k T~, \\
\widehat {\mathcal{R}} - {\mathcal{R}} &=& {\mathcal{R}} + \Hu T~, \\
\widehat {\mathcal{K}} - {\mathcal{K}}   &=& \frac{1}{\Hu}\left(\dot{\Hu} - \Hu^2 \right) T =
                         \frac{\dot{\overline{H}}}{\overline{H}} T~, \label{dHm}
\end{eqnarray} 
where $\overline{H}=\Hu/a$ is the usual Hubble parameter defined in the proper time. From the definition 
of Eq.~(\ref{dHm}) we see explicitly that we can identify ${\mathcal{K}} $ as the perturbation of $H$.

We now define gauge invariant quantities associated to the metric and fluid perturbations.  
Bardeen metric gauge invariants are defined \cite{Bardeen:1980kt} as:
\begin{eqnarray}
\Psi_B &=& A - \frac{\Hu}{k} \sigma_g - \frac{1}{k} \dot\sigma_g \label{eq:bardeenpsi}~,  \\
\Phi_B &=& H_L+\frac{1}{3}H_T - \frac{\Hu}{k} \sigma_g~.
\label{eq:bardeenphi} 
\end{eqnarray}
One can also build the following gauge invariant observable related to the expansion rate perturbation:
\begin{eqnarray}
{\mathcal{C}}  &=& {\mathcal{K}}  - \frac{1}{k} \frac{\dot{\bar H}}{\bar H} \sigma_g = 
{\mathcal{K}} - \frac{3}{2} (1+w_T) \frac{\sigma_g}{k\Hu}~, \label{eq:giXp} \\
{\mathcal{C}}_a  &=& {\mathcal{K}}_a  - \frac{1}{k} \frac{\dot{\bar H}}{\bar H} \sigma_g = 
{\mathcal{K}}_a - \frac{3}{2} (1+w_T) \frac{\sigma_g}{k\Hu}~.
\end{eqnarray}
It is also useful to define the following gauge--invariant quantity:
\begin{equation}
\mathcal{A} = \Psi_B - \frac{\dot{\Phi}_B}{\Hu} - \left( 1 - \frac{\dot\Hu}{\Hu^2} \right) \Phi_B
                    = \frac{3}{2}\left(1+w_T\right) \left(\widetilde{V}_T - \Phi_B \right)~, 
                    \label{eq:KSA}
\end{equation}
with $\widetilde{V}_T$ the (reduced) gauge invariant velocity of the total fluid defined by:
\begin{equation}
 (1+w_T) \widetilde{V}_T=\sum_a (1+w_a) \Omega_a \widetilde{V}_a~.
\label{eq:V_T}
\end{equation}


%
\subsection{Useful equations}
The perturbation equations for the metric can be derived from Einstein
equations: 
\begin{eqnarray}
\Phi_B + \Psi_B &=& - 3 \frac{\Hu^2}{k^2} \frac{p_T \Pi_T}{\rho_T}  
           =    - \frac{\Hu^2}{k^2} \Omega_\nu \Pi_\nu  = - \Omega_\nu \widetilde{\Pi}_\nu 
           \qquad \left(\widetilde{\Pi}=\frac{\Pi}{x^2}\right)  \label{pertEin0}~, \\
\Psi_B - \frac{\dot{\Phi}_B}{\Hu} &=&  \frac{3}{2}\frac{\Hu}{k} \left(1+w_T\right) V_T = 
            \frac{3}{2} \sum_a \left(1+w_a\right)\Omega_a \widetilde{V}_a \qquad 
             \left(\widetilde{V}=\frac{V}{x}\right)~, \label{perEin1} \\
\Phi_B &=&\frac{\Delta_T + 3 (1+w_T) \widetilde{V}_T}{3 (1+w_T) + \frac{2}{3}x^2 }=
                      \frac{\sum_a  \left(\Delta_a + 3\left(1+w_a\right) \widetilde{V}_a\right)\Omega_a}
                      {\sum_a 3\left(1+w_a\right) \Omega_a+ \frac{2}{3} x^2}~,  \label{perEin2}
\end{eqnarray}
where we have defined $x=k/\Hu$. From the previous equation one can
obtain the following relation for the expansion rate perturbations:
\begin{eqnarray}
\mathcal{C}  &=& \left[\frac{x^2}{3} - \frac{3}{2} \left(1+w_T\right)\right] \widetilde{V}_T \, ,\\
\mathcal{C}_a &=& \frac{x^2}{3} V_a - \frac{3}{2} \left(1+w_T\right) \widetilde{V}_T~.
\end{eqnarray}
%
%
For the sake of completeness we also provide the relation between the entropy perturbation
$\Gamma_a$, defined in Eq.~(\ref{KSPi}), and the sound speed in the rest frame of the fluid 
${c}^2_{Sa}$ which is given by:
\begin{equation}
  w_{a} \Gamma_{a}=\left({c}^2_{Sa}- c^2_{Aa}\right)\left[\Delta_{a}-
    \frac{\dot{\bar \rho}_{a}}{\bar \rho_{a}} \left(\frac{\Phi_B}{\cal H}-
    \frac{V_{a}}{k}\right)\right] \, .
\label{eq:Gammac2}
\end{equation}
%
\bibliographystyle{unsrt} 
\bibliography{bibdmde-v2.bib}
%
%
\end{document}